\documentclass[twocolumn,showpacs,preprintnumbers,amsmath,amssymb,prl]{revtex4}

\usepackage{graphicx}
\usepackage{dcolumn}
\usepackage{bm}
\usepackage{epsf,epsfig,latexsym}

\begin{document}
\title{Caloric Curves and Nuclear Expansion}

\author{J. B. Natowitz}
\author{K. Hagel}
\author{Y. Ma}
\author{M. Murray}
\author{L. Qin}
\author{S. Shlomo}
\author{R. Wada}
\author{J. Wang}
\affiliation{Cyclotron Institute, Texas A\&M University,\\
  College Station,  Texas, 77845}%

\date{\today}

\begin{abstract}
Nuclear caloric curves have been analyzed using an expanding Fermi gas 
hypothesis to extract average nuclear densities. In this approach the 
observed flattening of the caloric curves reflects progressively increasing 
expansion with increasing excitation energy. This expansion results in a 
corresponding decrease in the density and Fermi energy of the excited 
system. For nuclei of medium to heavy mass apparent densities 
$\sim0.4\rho_0$ are reached at the higher excitation energies.
\end{abstract}

\maketitle

In a recent paper measurements of nuclear specific heats from a 
large number of experiments were employed to construct caloric curves 
for five different regions of nuclear mass\cite{natowitz02}. Within 
experimental uncertainties each of these caloric curves exhibits a plateau 
region at higher excitation energy. In reference \onlinecite{natowitz02} 
it was pointed out that applying the simple Fermi gas relationship between 
excitation energy, $E^*$, and temperature, $T$, for a nucleus of mass ,$A$, 

\begin{eqnarray}
E^*=\frac{A}{K}T^2
\end{eqnarray}

\noindent where $K$ is the inverse nuclear level density parameter, to a 
large body of caloric curve data leads, at low excitation energy, to 
an increase in the apparent value of $K$,  followed 
at higher excitation energies by a rather rapid decrease in the apparent 
value of $K$\cite{natowitz02,hasse86,bortignon87,shlomo90,shlomo91,natowitz95}.
These results are presented in Figure 1. For this figure the data presented 
in Figure 5 of reference \onlinecite{natowitz02} have been smoothed by 
making a running average of  three points adjacent in excitation energy. 
This reduces the scatter in the points and makes the trends in the data 
clearer. Together these data span a range of fragmenting system masses 
$30<A<240$. 

Previous work has shown that the initial heating of the system leads to a 
small expansion of the nucleus, the washing out of shell 
effects and an increase in the nucleon effective 
mass\cite{hasse86,bortignon87,shlomo90,shlomo91,natowitz95}. This  
results in a decrease of the nuclear level density, reflected as an 
increase in the inverse level density parameter, 
$K$\cite{hasse86,bortignon87,shlomo90,shlomo91,natowitz95}. 
The magnitudes of these effects are mass dependent\cite{shlomo90,shlomo91}. 
These trends are seen very clearly in Figure 1 at excitation energies 
$<3$ MeV/nucleon. However, at excitation energies ranging from 
$\sim3$ MeV/nucleon for the heaviest systems to $\sim8$ MeV/nucleon for 
the lighter systems, the inverse level density parameters start to 
decrease with increasing excitation energy, reaching low values, 
near 6 MeV, at the highest excitation. (Exceptions are the three highest 
points in the A= 100-140 window. As discussed in 
reference \onlinecite{natowitz02}, these data points may indicate onset 
of a qualitatively different feature but this is not verified by the other 
data available.)                               

As can be seen in Figure 1, the general trend of the decrease of $K$ is 
similar to that which results from a temperature which remains constant 
with increasing excitation energy. This onset of the decrease, reflecting 
the onset of a plateau in the caloric curves has been 
taken as the signal of a new behavior. Based on comparisons with 
calculations\cite{hasse86,bortignon87,shlomo90,shlomo91,bonche84}
we  and others have previously described it as a departure from 
normal Fermi gas behavior.  
 
To explore the nature of that departure in greater detail we note that 
in the simple, non dissipative, uniform density Fermi gas 
model\cite{preston62} the relationship between the thermal 
excitation energy per nucleon, $\epsilon_{th}$ , the temperature, $T$, and the 
Fermi energy , $\epsilon_F$, is just

\begin{eqnarray}
\epsilon_{th}=\frac{\pi^2T^2}{4\epsilon_F(\rho)}
\end{eqnarray}
\noindent Using the usual expression for  a, the Fermi gas level density 
parameter, 
\begin{eqnarray}
a=\frac{A}{K(\rho)}=\frac{\pi^2}{4\epsilon_F(\rho)}
\end{eqnarray}

\noindent $K(\rho)$, the inverse level density parameter for an expanded 
nucleus of equilibrium density, $\rho_{eq}$, may be
 written\cite{norenberg02}

\begin{eqnarray}
K(\rho_{eq})=\frac{T^2}{\epsilon_{th}}\bigg(\frac{\rho_{eq}}{\rho_0}\bigg)^\frac{2}{3}
\bigg(\frac{m^*(\rho_0)}{m^*(\rho_{eq})}\bigg)
\end{eqnarray}
                                                              
where $\rho_0$ is the normal nuclear density and $m^*$ is the ratio of the 
effective mass of the nucleon to the mass of the free nucleon. At the 
temperatures at which we are interested in using this expression $m^*$ 
should be close to 1, indicating that the shell effects and collective 
effects leading  to increased level density above that predicted by the 
Fermi gas model are no longer 
important\cite{hasse86,bortignon87,shlomo90,shlomo91}.
Above the excitation energy where $m^*$ goes to 1, the relative density 
of a heated uniform Fermi gas at equilibrium  can be extracted from 
the ratio of the apparent inverse level density parameter, $K(\rho_{eq})$,
to $K_0$, the inverse level density parameter for a uniform Fermi gas,  
with $m^*$ assumed to be 1, at $T = 0$.

\begin{eqnarray}
\frac{\rho_{eq}}{\rho_0}=\bigg(\frac{K(\rho_{eq})}{K_0}\bigg)^\frac{3}{2}
\end{eqnarray}

We have used this relationship of equation (5) to derive the excitation 
energy dependence of the relative nuclear density, $\rho_{eq}/\rho_0$.  We 
applied this method to the determination of $\rho_{eq}/\rho_0$ points for 
each mass region only at excitation energies at and above the entry points 
into the caloric curve plateaus. These points have been derived from the 
caloric curves in reference \onlinecite{natowitz02} and shown to have a 
mass dependence consistent with that predicted by calculations which 
indicate the onset of Coulomb instabilities in the heated 
nucleus\cite{bonche84,levit85,besprovany89}. In evaluating this 
expression we have used the Fermi energy determinations of 
Moniz {\it et al.}\cite{moniz71} in equation (3) to determine the 
appropriate reference value of $K_0$ for the mass region being investigated.

To evaluate K($\rho_{eq}$) and the density it is required that 
$\epsilon_{th}$ be determined. We used an iterative technique to derive of 
$\rho_{eq}$/$\rho_0$ and $\epsilon_{th}$. This technique consisted of first 
estimating the relative density as (K$_{app}$/K$_0$)$^{3/2}$, employing this 
density estimate to evaluate the binding energy difference associated with 
expansion and subtracting that from the total excitation energy E$^*$ to 
obtain $\epsilon_{th}$.  Using this estimate of $\epsilon_{th}$ allowed a new 
estimate of  K($\rho_{eq}$) and density to be made. The process was 
repeated until self consistent values of $\epsilon_{th}$ and 
$\rho_{eq}/\rho_0$ were obtained.  The binding energies at different 
densities were calculated using a liquid drop model mass 
formula\cite{kleban02} with the bulk term extracted from the density and 
relative neutron excess dependent nuclear matter equation of state of 
Myers and Swiatecki\cite{myers98}. The surface term was scaled as the bulk
term and the Coulomb term corrected for the expansion. The binding energy 
changes evaluated in this way show a near parabolic dependence on density 
similar to that employed in the Expanding Emitting Source Model of 
Friedman\cite{friedman88}. However the absolute values are slightly lower.

The results, derived from the inverse level density data of Figure 1, 
are presented in Figure 2 as a plot of relative density, $\rho_{eq}/\rho_0$, 
versus excitation energy per nucleon. While the mass dependence seen in 
Figure 1 continues to be apparent in Figure 2, the general trend of 
decreasing nuclear density is similar in each mass region. In the present 
ansatz, the decreasing values of $K$ with excitation energy, seen in figure 1 
lead to decreasing nuclear densities. We see that at the entry into the 
caloric curve plateaus, densities near $0.7\rho_0$ are indicated. Over a 
span of a few MeV/nucleon in excitation energy the densities drop into the 
range of  $\sim0.4\rho_0$,  a value indicated by the thin horizontal line on 
the figure.  At higher excitation the densities extracted in this manner 
remain near $0.4\rho_0$.
                                    
Modest expansion is a general feature of calculations focusing on the 
thermal properties of heated nuclei\cite{shlomo91,bonche84,levit85,hasan98}. 
For comparison to the present results we have plotted in Figure 2 calculated 
average densities derived from the recent finite temperature Hartree-Fock 
(FTHF) calculations of Hasan and Vary for $^{90}$Zr\cite{hasan98} and 
densities for $A=120$ estimated from a model which assumes a trapezoidal 
density distribution and phenomenological dependencies on temperature 
derived from previous works\cite{shlomo91,shlomo02}. As 
seen the two results are quite similar. The result of the FTHF 
calculations are limited to energies below $\sim4$ MeV because the model space 
takes a limited number of oscillator shells in to account. Comparison of 
the data with these calculations indicates that above the Coulomb 
instability point the density decreases more rapidly with increasing 
excitation energy.

While direct measurement of the size or density of the fragmenting 
system are difficult to obtain, many previous theoretical and experimental 
results have suggested that low densities such as those indicated by 
the present analysis are reached in multi-fragmenting 
systems\cite{suraud89,gross93,bondorf85,friedman88,sugawa99,feldmeier00,%
gulminelli97,desesquelles98,schwarz01,cibor00}. 
We have previously reported average densities deduced from 
coalescence measurements for fragmenting medium mass nuclei\cite{cibor00}.
In Figure 3, the average densities obtained with that coalescence approach 
are compared with the present results for the $100<A<140$ mass region. 
The average densities indicated by these two independent techniques are 
seen to be in good agreement.  Bracken {\it et al.}\cite{bracken96}
have extracted sizes and average densities versus excitation energy by 
analyzing barriers for ejectiles in high energy $^3$He + $^{197}$Au 
collisions.  These, also plotted in figure 3, result in slightly lower 
densities.

In conclusion, it appears that this simple hypothesis of expansion of 
an equilibrated non-dissipative Fermi gas is sufficiently accurate to 
allow density information to be derived from reported caloric curve 
measurements. The present experimental results indicate that, 
as nuclei are heated, a more rapid rate of expansion with increasing 
excitation energy commences at the Coulomb instability temperature. 
This was suggested already in the early work of Bonche, Vautherin and 
Levit\cite{bonche84,levit85}. The mass dependence of the Coulomb 
instability temperature is such that the excitation energy and temperature 
at which this occurs increases with decreasing mass. At the higher 
excitation energies densities near $0.4\rho_0$ are reached. 
For medium to heavy mass systems, some recent studies have presented 
evidence for observation of critical behavior in the excitation 
energy range where the present work indicates a rapid decrease 
in density\cite{hauger00,elliott02,kleine02}. The critical 
temperatures reported in those studies are close to the Coulomb 
instability temperatures. The influence of the Coulomb energy on 
observables used to determine the critical point is not completely 
clarified at present\cite{srivastava01,richert01}.
 
\section{Acknowledgements}

This work was supported by the U S Department of energy under Grant DE-FG03-
93ER40773 and by the Robert A. Welch Foundation.

\begin{figure}[b]
\epsfig{file=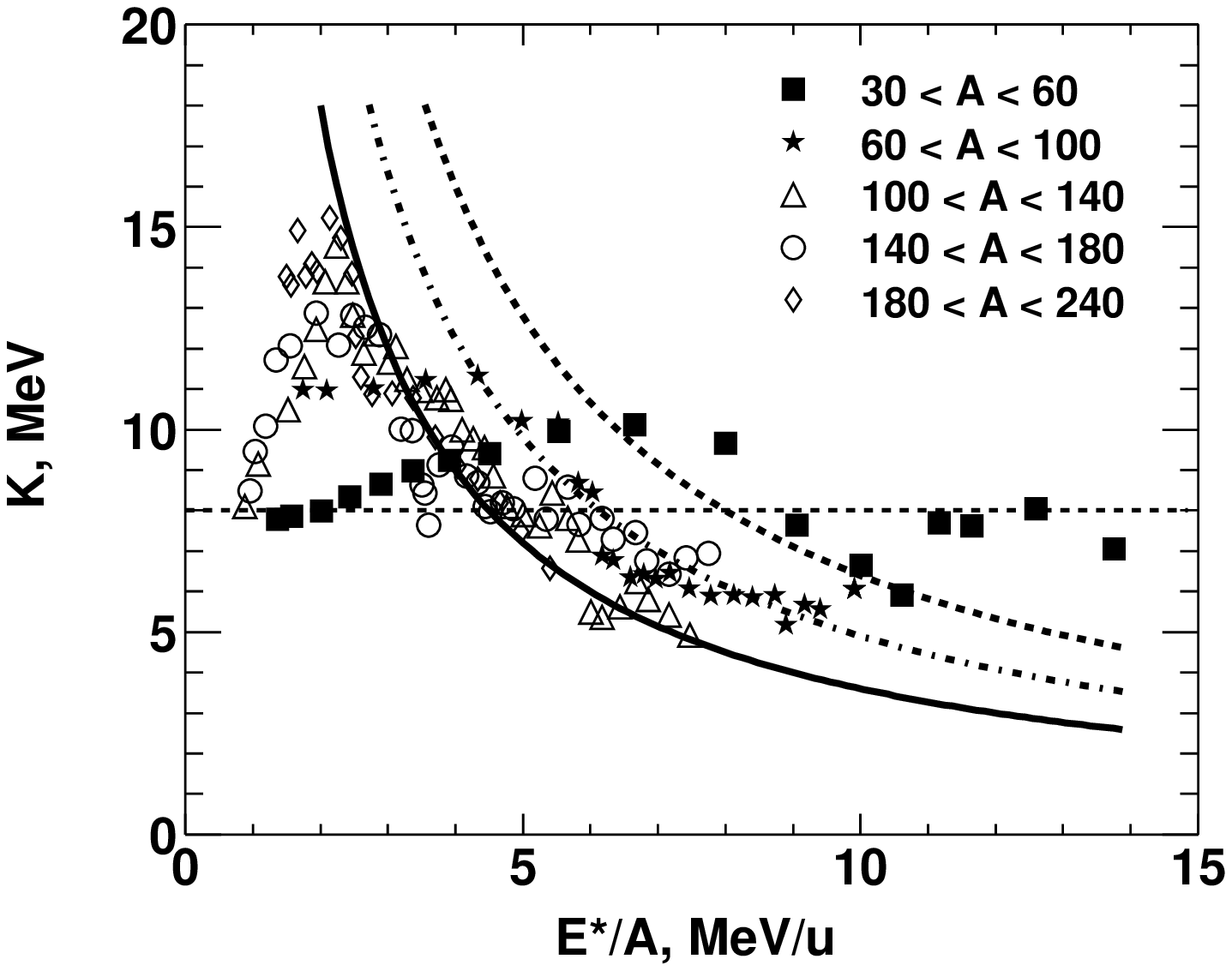,width=9.2cm,angle=0}
\caption{\label{fg:InverseLevelDensity}
Apparent inverse level density parameter K versus excitation energy.  
Symbols show results for five different mass regions; 
The meaning of the various symbols is indicated in the figure.
The thicker lines represent $K$ values calculated assuming constant 
temperatures of 6 MeV(solid), 7 MeV(dot-dashed) and 8 MeV (dashed).
}
\end{figure}
                           
\begin{figure}
\epsfig{file=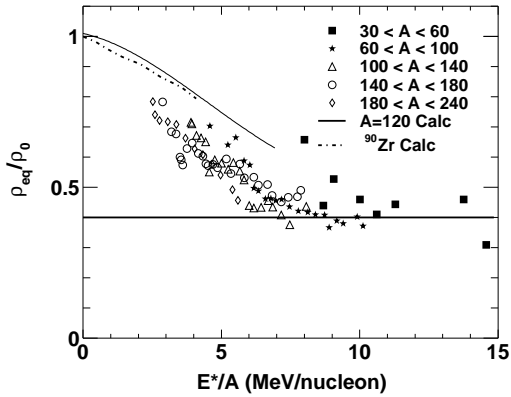,width=9.2cm,angle=0}
\caption{\label{fg:AverageDensity1}
Derived average nuclear density  versus excitation energy. 
Symbols show results for five different mass regions; 
The meaning of the various symbols is indicated in the figure.
The thin horizontal line marks $\rho = 0.4$. The two lines 
starting at $\rho= 1$ represent calculated average densities for 
$^{90}$Zr\cite{hasan98} (solid) and A=120\cite{shlomo91,shlomo02} 
(dot-dashed)(See text)
}
\end{figure}
 
\begin{figure}
\epsfig{file=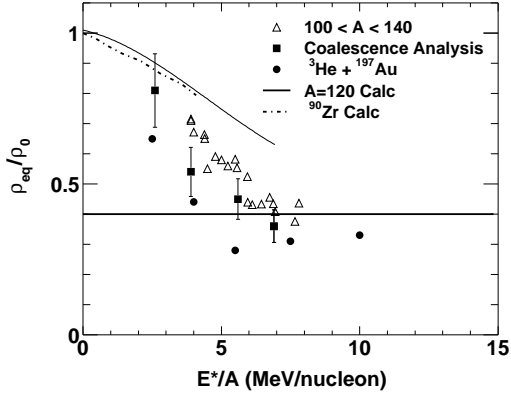,width=9.2cm,angle=0}
\caption{\label{fg:AverageDensity2}
Average nuclear density  versus excitation energy. Open diamonds
show results for $100<A<140$, derived in the present work. Solid squares
with error bars represent results of a coalescence model analysis of light
particle emission\cite{cibor00}. Solid circles represent results from
reference \onlinecite{bracken96}.  The thin horizontal line marks 
$\rho = 0.4$. The two lines starting at $\rho= 1$ represent calculated 
average densities for $^{90}$Zr\cite{hasan98} (dot-dashed) and 
$A=120$\cite{shlomo91,shlomo02} (solid) (See text)
}
\end{figure}
                           
\end{document}